\documentclass[a4paper,fleqn,usenatbib]{mnras}

\usepackage{newtxtext,newtxmath}

\usepackage[T1]{fontenc}
\usepackage{ae,aecompl}

\usepackage{graphicx}	
\usepackage{amsmath}	
\usepackage{amssymb}	

\title[Cosmological simulations and submillimetre surveys]{Comparison of cosmological simulations and deep submillimetre galaxy surveys}

\author[S. Aoyama et al.]{Shohei Aoyama,$^{1}$\thanks{E-mail: saoyama@asiaa.sinica.edu.tw (SA)}
Hiroyuki Hirashita,$^{1}$
Chen-Fatt Lim,$^{1,2}$
Yu-Yen Chang,$^{1}$\newauthor
Wei-Hao Wang,$^1$
Kentaro Nagamine,$^{4,5,6}$
Kuan-Chou Hou,$^{1,2,3}$
Ikkoh Shimizu,$^{4}$\newauthor
Hui-Hsuan Chung,$^{7}$
Chien-Hsiu Lee$^{8}$ and
Xian-Zhong Zheng$^{9}$
\\
$^{1}$Institute of Astronomy, and Astrophysics, Academia Sinica, Astronomy-Mathematics
Building, \\\hspace{0.3cm}AS/NTU No.\ 1, Sec.\ 4, Roosevelt Road, Taipei 10617, Taiwan\\
$^{2}$Department of Physics \& Institute of Astrophysics, National Taiwan University, Taipei 10617, Taiwan \\
$^{3}$Physics Department, Ben-Gurion University of the Negev, POB 653, Be'er Sheva 84105, Israel\\
$^{4}$Theoretical Astrophysics, Department of Earth \& Space Science, Osaka University, 1-1 Machikaneyama, Toyonaka, Osaka 560-0043, Japan\\
$^{5}$Department of Physics \& Astronomy, University of Nevada, Las Vegas, 4505 S. Maryland Pkwy, Las Vegas, NV 89154-4002, USA \\
$^{6}$Kavli IPMU (WPI), The University of Tokyo, 5-1-5 Kashiwanoha, Kashiwa, Chiba 277-8583, Japan \\
$^{7}$Institute of Astronomy, National Tsing Hua University, No. 101, Section 2, Kuang-Fu Road, Hsinchu 30013, Taiwan \\
$^{8}$National Optical Astronomy Observatory, 950 N Cherry Ave., Tucson, 85719 AZ, USA\\
$^{9}$Purple Mountain Observatory, Chinese Academy of Sciences, 8 Yuanhua Road, Nanjing 210034, China
}

\date{Accepted XXX. Received YYY; in original form ZZZ}

\pubyear{2018}

\begin{document}
\label{firstpage}
\pagerange{\pageref{firstpage}--\pageref{lastpage}}
\maketitle

\begin{abstract}
Recent progress in submillimetre surveys by single-dish telescopes allows us 
to further challenge the consistency between cosmological simulations and observations.  
In particular, we compare our simulations that include dust formation and destruction 
with the recent SCUBA-2 surveys (`STUDIES')
by putting emphases on basic observational properties of dust emission 
such as dust temperature, size of infrared (IR)-emitting region,
IR luminosity function and IRX--$\beta$ relation.
After confirming that our models reproduce the local
galaxy properties, we examine the STUDIES sample at $z\approx 1-4$, finding that
the simulation reproduces the aforementioned quantities
except for the $z\gtrsim 2$ IR luminosity function at the massive end
($\sim 10^{13}$\,L$_{\odot}$).
This means that the current simulation correctly reproduces the overall scaling 
between the size and luminosity (or star formation rate) of dusty region, 
but lacks extreme starburst phenomena at $z\gtrsim 2$.
We also discuss extinction curves and possible AGN contribution.
\end{abstract}




\section{Introduction}
Infrared (IR)\footnote{In this paper,  the term `IR' is used to refer to
the rest-frame wavelength range (including submillimetre)
where the emission is dominated by dust: $8\,\mu $m $<\lambda<1000\,\mu$m .} observations of galaxies are essential for 
studying galaxy evolution. 
Dust grains absorb stellar ultraviolet (UV)-to-optical radiation and
reemit it in the IR, thereby strongly modifying the 
spectral energy distribution (SED) in the UV--IR wavelength range. 
Therefore the dust emission in galaxies traces the obscured star formation activities, 
and it accounts for a dominant fraction of star formation activities
toward higher redshifts of $z\sim 2$, 
when the cosmic star formation rate (SFR) peaks in the
entire history of the Universe
\citep[e.g.][and references therein]{2005A&A...440L..17T,2013A&A...554A..70B}.
Thus, dust extinction and emission are of fundamental importance in deriving
the star formation history in the Universe.
A large number of IR-detected galaxies are barely visible in the optical and UV
\citep[][]{2013ApJ...769..116H,2013ApJ...775...61C}.

The luminosity function (LF) is one of the most important properties
that represent the statistical nature of galaxies. In particular,
the IR LF reflects
not only the number density of IR-luminous galaxies,  
but also the dust abundance in the Universe
and the nature of central engines of the luminous sources.
Statistical properties  of IR-luminous objects have been derived using,
for example, the \textit{Herschel} data.
\citet{2013MNRAS.432...23G} obtained 
not only the IR LF at $z=0$ but also its evolution up to
$z\sim 4$ \citep[see also][]{2013A&A...553A.132M}.
The typical detection limit of \textit{Herschel} survey is
$L_\mathrm{IR}\sim 10^{11.5}$\,L$_{\sun}$ at $z\sim 1$ and
$L_\mathrm{IR}\sim 10^{12.5}$\,L$_{\sun}$ at $z\sim 2$.

\cite{2013MNRAS.433..695C} combined \textit{Herschel}
data with {\it Wide-field Infrared Survey Explorer} (\textit{WISE}), 
{\it Spitzer}, and \textit{Infrared Astronomical Satellite} ({\it IRAS}) observations to 
investigate the properties of a flux-limited sample of local star-forming galaxies.
They fit the SEDs with modified blackbody spectra,
and showed that they are well-described by dust emission components with
emissivity indices $\beta\simeq 2$ and dust temperatures in
the range of 10--25 K.
They also showed that the dust temperature is strongly influenced
by the ratio of SFR to total dust mass of galaxies, indicating
that ongoing star formation activities are important for dust heating.
\cite{2013MNRAS.431.2317S} performed a comprehensive study of SEDs and
dust temperatures of IR-luminous galaxies at $ 0.1 < z < 2$. 
They used {\it Herschel} data from the deepest
Spectral and Photometric Imaging Receiver (SPIRE) and
Photodetecting Array Camera and Spectrometer (PACS) surveys in 
the Cosmological Evolution Survey (COSMOS),
the Great Observatories Origins Deep Survey (GOODS)-S and GOODS-N
fields, and examined the dust properties of
IR-luminous ($L_{\rm IR}>10^{10}$\,L$_{\odot}$) galaxies. 
They showed that the dust temperatures of galaxies with $L_{\rm IR}\sim 10^{12.5}$\,L$_{\odot}$
are significantly higher than those with $L_{\rm IR}\sim 10^{10.5}$\,L$_{\odot}$.

All the above studies have expanded our views on the IR galaxies at high redshift;
however, it is important to note that high-$z$ observations by \textit{Herschel} were strongly
affected by the confusion limit \citep[e.g.][]{2017MNRAS.471.4155K}.
To obtain a deeper view of dust emission in the Universe,
interferometric observations at submillimetre (submm) wavelengths are useful, 
e.g., by the Atacama Large Millimetre/submillimetre Array (ALMA), which has much higher
spatial resolution and sensitivity than \textit{Herschel}.
For example, \cite{2017ApJ...850...83F}, using the high-resolution capability of
ALMA, indeed constrained the size of dust-emitting region in individual
IR-luminous galaxies at $z=0$--6, and found that the IR-emitting region becomes
smaller at higher redshift although the dependence is weak.
Since the radiation field for a certain luminosity depends on the size of
the emitting region, spatially resolving the IR emission is important.
\cite{2018arXiv180203117C} argued theoretically that the surface brightness of
star-forming galaxy cannot exceed $\sim 10^{13}\,\mathrm{L}_{\odot}\,{\rm kpc}^{-2}$,
which is determined by the Eddington-limited star formation activity.
With the size of dust-emitting region
($\sim 1$\,kpc) obtained for IR-luminous objects by \cite{2017ApJ...850...83F},  
extremely IR-luminous objects ($L_{\rm IR}>10^{13}$ L$_{\odot}$) might host a luminosity
source other than stars, most probably active galactic nuclei (AGNs).

High spatial resolution is not the only important factor.
Because of their narrow field of view, interferometers are generally not suitable
for a wide-field survey.
Deep surveys with a large ground-based single-dish telescope is a viable way of
deriving statistical properties of galaxies up to high redshift. 
For example, the James Clerk Maxwell Telescope (JCMT)
[the Submillimetre Common-User Bolometer Array (SCUBA) on JCMT]
was used to investigate the dust properties at various redshifts.
\cite{2005MNRAS.364.1253V} derived the local submm luminosity and 
dust mass functions by the SCUBA Local Universe Galaxy Survey (SLUGS) and 
the \textit{IRAS} Point Source Catalog Redshift Survey (PSCz).
The submm observations are suitable for tracing the cold
dust component (17--24 K in their sample for $\beta =2$),
which usually dominates the total dust mass in galaxies.
They showed that, after an appropriate treatment of IR SEDs, the
dust mass functions derived from those two surveys agree well.

In general, the IR SED depends not only
on the total dust amount but also on the dust temperature.
The dust temperature reflects the stellar radiation field incident on the
dust; thus, it provides us with important information on the relative spatial distribution
of dust to stars. In addition, if we analyze UV attenuation properties, especially
based on the UV spectral slope $\beta_{\rm UV}$,\footnote{This is usually denoted as
$\beta$, but in order to avoid confusion with the IR emissivity index,
we denote the UV SED slope as $\beta_\mathrm{UV}$.}
we are able to infer how the
reddening at UV wavelengths affects the IR luminosity \citep{1999ApJ...521...64M}.
 Therefore, a deep survey that satisfies the following conditions
is desirable: (i) it covers a large area; (ii) the area is observed at two IR wavelengths
(one of the wavelengths should be near the SED peak)
so that the derivation of dust temperature is possible; and (iii) the detected galaxies
are followed up at rest-frame UV wavelengths.
The current deepest survey that satisfies all the above (i)--(iii) is
the SCUBA-2 Ultra Deep Imaging EAO Survey (STUDIES) as we introduce below.

The SCUBA-2 camera on JCMT is a powerful instrument for
surveying and characterizing the submm emission in the distant Universe with
a high simultaneous mapping capability at 450 and 850\,$\micron$.
STUDIES realized both substantial sensitivity and observing area
\citep{2017ApJ...850...37W}. 
The program has been allocated  {650} hours on SCUBA-2 in order to obtain 
confusion-limited 450 and 850 $\mu$m images in the Cosmic Assembly Near-infrared
Deep Extragalactic Legacy Survey (CANDELS) regions
in the COSMOS and Subaru/XMM-Newton Deep Survey (SXDS) fields.
The program is still ongoing,
and it has already produced one of the deepest 450 $\micron$ maps
in the COSMOS-CANDELS region.
Before STUDIES, the population of known submm galaxies (SMGs) were
limited to ultra-luminous ($L_{\rm IR}\gtrsim 10^{12}$\,L$_{\odot}$) ones for $z>1$ 
by {\it Herschel} \citep[e.g.][]{2013MNRAS.432...23G}.
By taking advantage of the fast mapping speed of SCUBA-2  {and the higher resolution at 450 $\micron$}, 
they claimed that the dust mass function can be derived down to the objects
ten times fainter than those detected by {\it Herschel}.
\cite{2017ApJ...850...37W} showed that the integrated surface brightness from 
their counts down to 1\,mJy is $90\pm 17.2\,{\rm Jy}\,{\rm deg}^{-2}$, 
which can account for up to $83^{+15}_{-16}$ \% of the {\it COBE} 450 $\mu$m background. \cite{2018arXiv180807480C} further showed that the survey can detect
$L_{\rm IR} < 10^{11}\, {\rm L}_{\odot}$ main-sequence star-forming galaxies up to $z \sim 3$.

Together with the observational advancement, 
there have also been significant theoretical developments in understanding the dust properties in
a cosmological volume.
\cite{2010MNRAS.403..620D} implemented models for dust formation by supernovae (SNe) and dust destruction by
SN shocks in their cosmological simulation to predict the detectability of dust emission
at $z\gtrsim 5$ by ALMA.
\citet{2017MNRAS.471.3152P} derived the time evolution of the cosmic dust abundance,
the dust mass function and the relation between dust-to-gas mass ratio and stellar mass 
with a semi-analytic model. \citet{2017MNRAS.468.1505M} treated dust as a component
associated with gas (treated as a fluid) in their {\small AREPO} cosmological hydrodynamical simulations. They broadly reproduced the cosmic dust abundance, the relation between SFR and stellar mass, the relation between dust-to-gas mass ratio and metallicity, and the dust mass function of SMGs.
\citet[][hereafter A18]{2018arXiv180204027A} performed {\small GADGET-3} cosmological hydrodynamic simulations with dust evolution, obtaining cosmic dust abundances at various redshifts consistently with observations. 
In their simulation, the chemical evolution code {\small CELib} \citep[][]{2017AJ....153...85S}
was coupled with the models of star formation, SN feedback, metal enrichment, 
dust formation and its destruction. 
The dust grain size distribution was treated by the `two-size approximation', in which 
large and small grains are separated at $\sim 0.03\,\mu$m \citep{2015MNRAS.447.2937H}.
This two-size approximation has also been applied to a simulation of galaxy clusters by 
\citet{2018arXiv180406855G}, who confirmed the importance of dust processing in the gas.
\citet{2018MNRAS.tmp.1185M} followed up their earlier work with an implementation of the evolution of grain size distribution in isolated galaxy simulations, providing some test calculations.
Thus, for cosmological simulations, the two-size approximation has been the most advanced
way of implementing the grain size information so far.

Since the extinction properties of UV stellar light strongly depend on the grain size
distribution \citep{2014MNRAS.440..134A},
A18's simulation based on the two-size approximation provides a unique tool that predicts the extinction curves in individual galaxies in a cosmological volume. 
Therefore, among the simulations that explicitly include dust evolution, A18's work is
the most suitable for the comparison with recent deep submm surveys. The only
caveat is the poor spatial resolution, which is unavoidable for a cosmological simulation that attempts to simulate a large sample of galaxies simultaneously. 
However, as shown later, we resolve the observationally suggested size of IR-emitting region
$\sim 1$\,kpc \citep{2017ApJ...850...83F} at high redshift. 
We also test our models against the observational data at $z=0$, and confirm that our simulation reproduces the observed properties of dust in the local Universe.
Thus, the aim of this paper is to examine the consistency between the state-of-the-art
large-volume simulation of dust enrichment and the deepest observation
of statistical dusty-galaxy properties.

In addition, taking advantage of the implementation of grain size distribution in our
simulations and  the existence of rest-UV
follow-ups for the observational data, we consider the so-called
IRX--$\beta_\mathrm{UV}$ relation.
The relation between IR excess (IRX: UV-to-IR luminosity ratio) and
$\beta_{\rm UV}$ was originally obtained by \cite{1999ApJ...521...64M}
for nearby starburst galaxies.  Theoretically,
$\beta_{\rm UV}$ can be predicted by a stellar population synthesis model
with dust extinction included.
The IRX--$\beta_\mathrm{UV}$ relation is observationally useful because it
enables us to infer what fraction of star formation activity is attenuated by dust
\citep[e.g.][]{2011ApJ...741..124H,2012ApJ...744..154R,2012ApJ...756...14S,2012ApJ...756..164F,2014ApJ...793..115B}.
The IRX--$\beta_\mathrm{UV}$ relation is still being debated.
\cite{2012ApJ...755..144T} reconsidered the IRX--$\beta_{\rm UV}$ relation
by correcting it for the photometric aperture effect, and obtained a `shifted'
IRX--$\beta_\mathrm{UV}$ relation relative to that derived by  \cite{1999ApJ...521...64M}.
There is also an indication that the IRX--$\beta_\mathrm{UV}$ relation is
different at $z\gtrsim 5$  \citep[][]{2015Natur.522..455C,2016ApJ...833...72B}.
Some theoretical models of IRX--$\beta_\mathrm{UV}$ relation show that
the relation is affected by the contamination of old stars and the
relative distribution of dust and stars (or radiation transfer effects) 
\citep[][]{2016MNRAS.462.3130M,2017MNRAS.471.5018F,2018MNRAS.474.1718N,2017MNRAS.472.2315P}.

This paper is organized as follows. We introduce the simulation and the analysis method
in Section \ref{sec:model}.
We test our results against observational data at $z=0$ in Section \ref{sec:local}.
In Section \ref{sec:STUDIES}, we compare the simulation results at $z=1$--4 with the STUDIES
galaxy survey data.
In Section \ref{sec:discussion}, we further discuss this comparison, including the uncertainties and limitations. 
Section \ref{sec:conclusion} concludes this paper.
We adopt the following cosmological parameters 
\citep{2016A&A...594A..13P}:
baryon density parameter $\Omega_{\rm b} = 0.049$, 
total matter density parameter $\Omega_{\rm m} =0.32$, 
cosmological constant parameter $\Omega_\Lambda =0.68$, 
Hubble constant $H_{0} = 67$ km s$^{-1}$ Mpc$^{-1}$, 
power spectrum index $n_{\rm s}=0.9645$, and
density fluctuation normalisation $\sigma_{8}=0.831$.
In this paper, we also use $h \equiv H_{0} / (100$ km s$^{-1}$ Mpc$^{-1})=0.67$
for the non-dimensional Hubble constant.

\section{MODEL}\label{sec:model}
\subsection{Cosmological simulation with dust evolution}

\begin{table}
\centering
\begin{minipage}{90mm}
\caption{Simulation Parameters}
\label{table:simulation}
    \begin{tabular}{cccccc}\\ \hline 
Name & Boxsize & $N$ & $\varepsilon_{\rm grav} $ & $m_{\rm dm}$ & $m_{\rm gas}^{\rm init}$ \\ 
     & [$h^{-1}$Mpc] &&[$h^{-1}$kpc]  &[$h^{-1}{\rm M}_{\odot}$]&[$h^{-1}{\rm M}_{\odot}$] \\ \hline 
L50N512 & 50 & $2\times 512^{3}$ & 3 &$6.89 \times 10^{7}$& $1.28 \times 10^{7}$\\ \hline 
L25N512 & 25 & $2\times 512^{3}$ & 1.6 &$8.61 \times 10^{6}$& $1.60 \times 10^{6}$ \\ \hline
    \end{tabular}
    \medskip\\
\textit{Note}: $N$, $\varepsilon_{\rm grav} $, $m_{\rm dm}$ and $m_{\rm gas}^{\rm init}$ are 
the total number of particles, the gravitational softening length, the mass of dark matter particle
and the initial mass of gas particle, respectively.
\end{minipage}
\end{table}

We use and post-process the simulation of A18 to derive the statistical
dust emission properties in a cosmological volume. We briefly review the
simulation here.
We performed cosmological $N$-body/smoothed particle hydrodynamic (SPH)
simulations with \textsc{gadget3-osaka}
developed by \cite{2017MNRAS.466..105A} and Shimizu et al.\ (2019, in preparation)
based on the \textsc{gadget-3} code 
\citep[originally described in][]{2005MNRAS.364.1105S}.  
Important parameters for the simulation setup are summarized in Table~\ref{table:simulation}, and L50N512 is our default simulation.

In addition to the L50N512 run in A18, we also performed another
simulation, referred to as L25N512, with a smaller box size in
order to test the effect of spatial resolution (Table~\ref{table:simulation}).
However, the number of
galaxies whose luminosities are in the range of the adopted observational sample
decreases because of the 8 times smaller volume.
Therefore, L25N512 is not very suitable for our main comparison with the observations,
and is only used to examine the effect of spatial resolution.
We also find that the effective resolution of L50N512
(sub-kpc) is sufficient for resolving the IR-luminous regions
(Section~\ref{subsec:Rdust-LIR}), but twice worse spatial resolution would fail to resolve them.
Note that we allow the SPH smoothing length to become as small as one-tenth of 
the gravitational softening length.  For example, the minimum SPH smoothing 
in the L50N512 run has decreased to 0.84\,kpc and 0.3\,kpc (both physical) for $z=0$ and $z=3$, respectively. 
Therefore, considering the currently available computational resource,
L50N512 is the optimum setup in terms of box size and resolution.
In other words, we are able to clarify how much such an
`optimum' simulation could reproduce the observational properties of IR-selected
galaxies, and extract physical information through the comparison with observations.

{ In order to clarify the resolution effects of simulations, A18 also
compared dust mass functions between L50N512 and a lower resolution simulation with $2 \times 256^{3}$ particles. There were no significant differences between the two results. We also confirmed that the basic relations between the dust abundance and other physical quantities do not significantly change between the low- and high-resolution runs. Thus, the convergence of dust properties
has already been confirmed.}
In the following, we display the results from L50N512 run unless otherwise stated.

We treat star formation, SN explosion, stellar feedback,
and metal enrichment by SNe and asymptotic giant branch stars consistently
using {\small CELib} \citep{2017AJ....153...85S}.
We solve the dust evolution on each gas particle.
In order to calculate the grain size distribution within a reasonable computational cost,
we adopt the two-size approximation, in which
the entire grain size range is represented by two sizes 
divided at grain radius $a\sim 0.03\,\mu$m \citep{2015MNRAS.447.2937H}.
The following processes are included for dust evolution:
dust condensation in stellar ejecta, dust growth in the dense and cold gas
via accretion and coagulation, dust disruption (or fragmentation) by
shattering in the diffuse medium,
dust destruction by sputtering in SN shocks and hot inter-galactic medium (IGM).
The simulation is not capable of resolving dense clouds 
in which dust grains grow via accretion and coagulation.
Thus, A18 adopted a subgrid model where 
each gas particle satisfying the `dense-gas condition' (number density $n_\mathrm{gas}>0.1$ cm$^{-3}$
and temperature $T_\mathrm{gas}<10^4$ K) hosts
dense clouds with $n_\mathrm{gas}=10^{3}$\,cm$^{-3}$ and $T_\mathrm{gas}=50$\,K
(the mass fraction of the dense clouds in the gas particle is assumed to be
10 per cent).
For stellar dust enrichment, we assume that 10 per cent of the metals
condense into dust. For the other processes, we adopt proper time-scales
for large and small grains 
basically estimated from the collision time-scales between dust grains or
between dust and gas, depending on the process. 
These time-scales generally depend on gas density, temperature, and metallicity
(see Sections 2.2--2.4 of A18 for detailed descriptions of each process).

Galaxies are identified using \textsc{P-Star groupfinder} 
\citep{2001MNRAS.328..726S}. In short, we find
the baryonic (gas + stars) density peaks in the smoothed density field
and measure the densities of $N_{\rm ngb}$ nearest neighbour particles
around the density peak.
We adopt $N_{\rm ngb}=512$ and $N_{\rm ngb}=128$
for L50N256 and L25N512, respectively.
The particle at the density peak is considered 
as a `head particle' around which a galaxy is identified if all the neighbor particles
have lower densities.

The most basic quantities in individual galaxies are the masses of dust and stars, 
which are needed to treat the absorption and reprocessed of stellar light by dust.
The spatial extent of dust and stars is also important since the heating of dust 
by the radiation field depends on how compactly dust and stars are distributed.
We find that the radial profile of dust mass density in a galaxy is
approximately described by an exponential function in our simulation, 
which provides the scale length $R_{\rm dust}$. 
In order to estimate $R_\mathrm{dust}$, we extract the (20 comoving kpc)$^{3}$
cubic region around each galaxy centre.
We cut each region into $20^{3}$ cubic grid. 
Thus the effective resolution of the radial profile estimate 
is 1\,kpc, which is comparable to the spatial resolution of our simulation.
By fitting the profile to an exponential function with the least-square method,
we obtain the scale length $R_\mathrm{dust}$, and adopt it as the radius of dusty region. 

To estimate the optical depth of dust, we first estimate the surface densities of
large and small grains, denoted as $\Sigma_{\rm L}$ and $\Sigma_{\rm S}$, respectively,
in the following manner:
\begin{eqnarray} 
M_\mathrm{L/S} & = & \int_0^{\alpha R_\mathrm{dust}}4\pi R^2\rho_\mathrm{L/S}\,\mathrm{d}R,\\
\Sigma_{\rm L/S}&=&\dfrac{M_{\rm L/S}}{\pi (\alpha R_{\rm dust})^{2}}~,\label{eq:surface_dust}
\end{eqnarray}
where 
$M_\mathrm{L}$, $M_\mathrm{S}$, $\rho_\mathrm{L}$, and $\rho_\mathrm{S}$
are the total mass of large grains, the total mass of small grains, the mass density of large
grains, and the mass density of small grains, respectively. 
The parameter $\alpha$ determines the radial range where the total dust mass is derived. We later fix the value of $\alpha$ in
Section \ref{subsec:Rdust-LIR}.
The total dust mass, $M_\mathrm{dust}$, is defined as
\begin{eqnarray}
M_\mathrm{dust}=M_\mathrm{L}+M_\mathrm{S}.
\end{eqnarray}

\subsection{Estimation of dust extinction optical depth}
Dust extinction including its wavelength dependence
(extinction curve) is estimated based on the
surface densities of large and small grains in equation (\ref{eq:surface_dust}).
For the calculation of extinction curve, we need a continuous functional
form of grain size distribution while we represent 
the entire grain size range by two sizes.
Following \cite{2015MNRAS.447.2937H} and \cite{2017MNRAS.469..870H},
we adopt the `modified-lognormal function' of grain size distribution for
large and small grains:
\begin{eqnarray} 
N_{i}(a)&=&\dfrac{C_{i}}{a^{4}}\exp\left( -\dfrac{\left(\ln (a/a_{0,i}) \right)^{2}}{2\sigma^{2}} \right)~,
\end{eqnarray}
where $a$ is the grain radius (we assume grains to be spherical),
$i=\mathrm{L}$ or S is the label for large or small grains, 
$C_{i}$ is the normalization constant, 
$a_{0,i}$ is the typical grain radius of each size domain
(we adopt $a_{0, {\rm L}}=0.1\,\mu {\rm m}$ 
and $a_{0, {\rm S}}=0.005\,\mu {\rm m}$), and
$\sigma$ is the standard deviation of the lognormal distribution.
The normalization $C_{i}$ is determined in a consistent manner with the total
surface densities estimated in Eq.~(\ref{eq:surface_dust}) as
\begin{eqnarray} 
\Sigma_{i}&=&\displaystyle\int_{0}^{\infty}\dfrac{4}{3}\pi a^{3}sN_{i}(a)\,\mathrm{d}a,
\end{eqnarray}
where $s$ is the material
density of dust grains. We adopt $s=3$ g cm$^{-3}$.

We calculate the optical depth $\tau_{\lambda}$ and 
the extinction curve $A_{\lambda}$ (magnitude of extinction as a function
of wavelength) as follows:
\begin{eqnarray} 
\tau_{\lambda}&=&\displaystyle\sum_{i={\rm L,S}}\displaystyle\int_{0}^{\infty} N_{i}(a)\pi a^{2}Q_{\rm ext}(a, \lambda)\,\mathrm{d}a~,\label{eq:tau}\\
A_{\lambda}&=& (2.5\log_{10}\mathrm{e})\tau_{\lambda}~,
\end{eqnarray}
where $Q_{\rm ext}(a, \lambda)$ is the extinction cross-section relative to the
geometric cross-section, which is evaluated
by using the Mie theory \citep[][]{1983asls.book.....B}.
For the optical constants necessary for the Mie theory calculation, we
need to specify the grain materials. However, in our simulation, we do not separate
grain species. Thus, we simply assume that all grains are composed of silicate, for
which we adopt the same optical constants as in \cite{2001ApJ...548..296W}.
To avoid the complexity in the prediction arising from the 2175 \AA\ bump,
we neglect graphite. Neglecting graphite empirically gives a good fit to the Small
Magellanic Cloud (SMC)
extinction curve \citep[][]{1992ApJ...395..130P,2001ApJ...548..296W,2016PASJ...68...94H}.
In other words, our model assumes a bumpless extinction curve, 
which will explain the observational data well, as we show later.
We emphasize that the steepness of extinction curve can still be estimated in
a manner consistent with the calculated grain size distribution.
We discuss the resulting extinction curves in Section \ref{subsec:var_ext}.

\subsection{Stellar SED and dust emission}

We now calculate the stellar SED for each galaxy taking into account the
extinction optical depth estimated in equation (\ref{eq:tau}).
We adopt a spectral synthesis model by \citet[][]{2003MNRAS.344.1000B}
to obtain the intrinsic SED of single stellar population at various ages.
The SED per stellar mass of single stellar population is denoted as
$\tilde{L}_{\lambda}$, which depends on the
stellar age ($t_*$) and metallicity ($Z_*$). We adopt the Chabrier initial mass function
(IMF) \citep[][]{2003PASP..115..763C}.
The intrinsic stellar SED, $L_{\lambda}^0$, is calculated by summing up all the
contribution from the stellar particles within the radius $R_\mathrm{dust}$ as
($k$ is the index of stellar particle)
\begin{eqnarray} 
L_{\lambda}^{0}&=&\displaystyle\sum_{\substack{k \\ 0\le R_{k} \le \alpha R_{\rm dust}}}\!\!\!\!\!
\tilde{L}_{\lambda}( t_{* (k)},\, Z_{* (k)})m_{{\rm star}(k)},
\end{eqnarray}
where $m_{{\rm star}(k)}$ and $R_{k}$ are the mass and the distance from the
galaxy centre for the $k$-th stellar particle.
The observed SED after dust extinction is calculated by
\begin{eqnarray}
L_\lambda =\left\{
\begin{array}{ll}
L_\lambda^0{\displaystyle \frac{1-\exp (-\tau_\lambda )}{\tau_\lambda}} & \mbox{(for $\lambda >912$\,\AA)},\\
0 & \mbox{(for $\lambda\leq 912$\,\AA)},\\
\end{array}
\right.
\end{eqnarray}
where we assume a mixed geometry between dust and stars
(which gives a better approximation for the dust--star geometry
within $\alpha R_\mathrm{dust}$ than a screen geometry) and
complete absorption of ionizing photons ($\lambda\leq 912$\,\AA) by hydrogen.
We use $\tau_\lambda$ calculated in equation (\ref{eq:tau}).
We assume that the extinguished light is reprocessed into the IR regime:
\begin{eqnarray} 
L_{{\rm IR}}&=&\displaystyle\int_{912\,\text{\AA}}^{\infty}(L_{\lambda}^0-L_\lambda )
\,\mathrm{d}\lambda~,\label{eq:reprocessing}
\end{eqnarray}
where $L_\mathrm{IR}$ is the total IR luminosity, which is directly compared with
the observationally derived IR luminosity at rest-frame $\lambda =8$--1000 $\micron$
by assuming that the dust luminosity emitted at wavelengths shorter than
8 $\micron$ or longer than 1000 $\micron$ is negligible.

\subsection{Derivation of dust temperature}
For simplicity, we assume that the dust emission follows the
so-called modified blackbody spectrum with a single dust temperature
$T_\mathrm{dust}$ (i.e.\ we neglect the variation of dust temperature within
a galaxy). With this simplification,
the IR SED $L_{\nu}^{\rm IR}$ is described by the following equations:
\begin{eqnarray} 
L_{\nu}^{\rm IR}&=&4\pi M_{\rm dust}\kappa_{\nu}B_{\nu}(T_{\rm dust})~,\\
\kappa_\nu & = & \kappa_0\left(\frac{\nu}{\nu_0}\right)^\beta ,
\end{eqnarray}
where $B_{\nu}(T)$ is the Planck function, $\kappa_{\nu}$ is
the mass absorption coefficient
at frequency $\nu$, $T_\mathrm{dust}$ is the dust temperature, and
$\kappa_0$ is the value of $\kappa_\nu$ at frequency $\nu_0$.
We adopt $\nu_{0}=3.00\times 10^{12}$\,Hz and $\kappa_{0}=52.2$\,cm$^{2}$\,g$^{-1}$ \citep{2010MNRAS.403..620D}.
Since $T_\mathrm{dust}\propto\kappa_0^{-1/(4+\beta)}$ under a fixed $L_\mathrm{IR}$,
the resulting dust temperature is not sensitive to the adopted value of $\kappa_0$.
With the above SED shape, the total IR luminosity $L_{\rm IR}$ is evaluated as
\citep[e.g.][]{2010MNRAS.403..620D}
\begin{eqnarray} 
L_{\rm IR}&=&\displaystyle\int_{0}^{\infty}L_{\nu}^{\rm IR}\mathrm{d}\nu\nonumber\\
&=&4\pi M_{\rm dust}\kappa_{\nu_{0}}\nu_{0}^{-\beta}
\left(\dfrac{k_{\rm B}T_{\rm dust}}{h} \right)^{4+\beta}\left(\dfrac{2h}{c^{2}} \right)\Gamma(4+\beta)\zeta(\beta+4)~,\label{eq:IR2}\notag \\ \label{eq:integ_LIR}
\end{eqnarray}
where $\Gamma(x)$ and $\zeta(x)$ are the Gamma function and the Riemann zeta function, respectively,  and $k_{\rm B}$, $h$ and $c$ are the Boltzmann constant, the Planck constant and the speed of light, respectively.
We assume radiative equilibrium for dust grains in order to derive the dust temperature,
so that we substitute equation~(\ref{eq:reprocessing}) for $L_\mathrm{IR}$ in
equation (\ref{eq:integ_LIR}).
By solving equation \eqref{eq:IR2} for $T_{\rm dust}$, we obtain
\citep[][]{2012MNRAS.427.2866S}
\begin{eqnarray} 
T_{\rm dust}&=&\dfrac{h}{k_{\rm B}}\left(
\dfrac{c^{2}\nu_{0}^{\beta}}{8\pi h \kappa_{0} \Gamma(4+\beta)\zeta(4+\beta)} 
\dfrac{L_{\rm IR}}{M_{\rm dust}}\right)^{\dfrac{1}{4+\beta}}.~ 
\label{eq:IR3}
\end{eqnarray}
We adopt $\beta =2$ unless otherwise noted \citep{2013MNRAS.433..695C}. 
In this case, equation~\eqref{eq:IR3} is reduced to\footnote{We note that
the numerical coefficient in equation~\eqref{eq:IR4} was given incorrectly
by \cite{2010MNRAS.403..620D}, and we correct it here. 
\citet{2012MNRAS.427.2866S} derived this coefficient as 7.64, and we confirm that we can reproduce their value using the dust size and $Q$-value in their paper.} 
\begin{eqnarray} 
T_{\rm dust}&=&
7.5 \left(\dfrac{L_{\rm IR}/L_{\odot}}{M_{\rm dust}/M_{\odot}}\right)^{\dfrac{1}{6}}~{\rm K}~.\label{eq:IR4}
\end{eqnarray}

\subsection{IRX-$\beta_{\rm UV}$ relation}

The IRX is defined as
\begin{eqnarray} 
{\rm IRX}&\equiv &\dfrac{L_{\rm IR}}{L_\mathrm{UV}(1650 \AA)},
\end{eqnarray}
where $L_\mathrm{UV}(\lambda )\equiv \lambda L_{\lambda}$ and
we adopt $\lambda =1650$ \AA\ following the UV wavelength 
of the STUDIES sample (see Section~\ref{subsec:sample}).
The UV SED slope,
$\beta_{\rm UV}$, which is defined by the relation
$L_{\lambda}\propto\lambda^{\beta_{\rm UV}}$ at rest-UV wavelengths,
is estimated as
\begin{eqnarray} 
\beta_{{\rm UV}}&=&\dfrac{\log_{10}(L_{\lambda_{2}})-\log_{10}(L_{\lambda_{1}})}{\log_{10}(\lambda_{2})-\log_{10}(\lambda_{1})}~,\label{def_of_beta}
\end{eqnarray}
where $\lambda_{1}$ and $\lambda_{2}$ are fixed to 1650 {\text \AA} and 2300 {\text \AA}, respectively, for the comparison with the STUDIES sample.
\begin{figure}
	\includegraphics[width=\columnwidth]{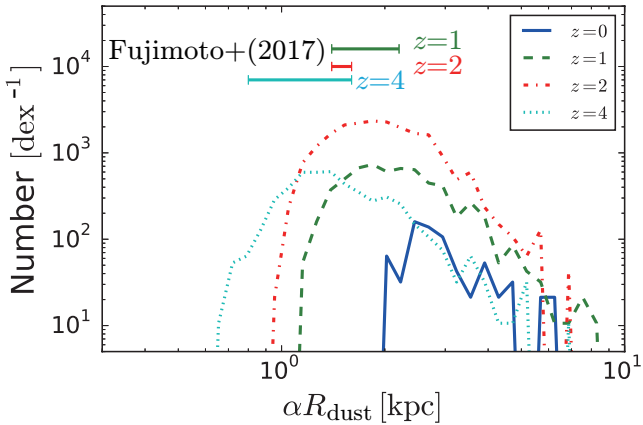}
    \caption{Histogram of the radius ($\alpha R_\mathrm{dust}$) of IR-emitting region at $z=0$, 1, 2, and 4 (solid, dashed, dot-dashed, and dotted lines, respectively) for the galaxies in the simulation box. We adopt $\alpha =2$ in this paper; that is, the IR-emitting region is twice the scale length of the dust distribution. We selected galaxies whose IR luminosity is greater than $10^{11}{\rm L}_{\odot}$. The horizontal bars show the corresponding observed ranges at $z=1$, 2, and 4 taken from \citet[][]{2017ApJ...850...83F}.  We find that $\alpha =2$ produces the radii consistent with the observed results including the evolution along the redshift. 
    }
    \label{fig:R_IR0}
\end{figure}

\section{Calibration and Tests}\label{sec:local}

Before comparison with the STUDIES sample, we check if our models
reproduce the principal properties of dust emission at $z=0$.
In the above, the radius of the IR-emitting regions relative to the scale length
of dust distribution ($\alpha$ in equation~\ref{eq:surface_dust}) was left undetermined.
We first determine it in Section \ref{subsec:Rdust-LIR}.
For the statistical properties of dust emission, we focus on the
IR LF. We also examine the $T_{\rm dust}$--$L_{\rm IR}$,
$R_{\rm dust}$--$L_{\rm IR}$ and
IRX--$\beta_{\rm UV}$ relations.
The dust mass and its relation to the metallicity have already been tested
by A18; thus, we concentrate on the IR emission properties in this paper.

\begin{figure*}
	\includegraphics[width=\textwidth ]{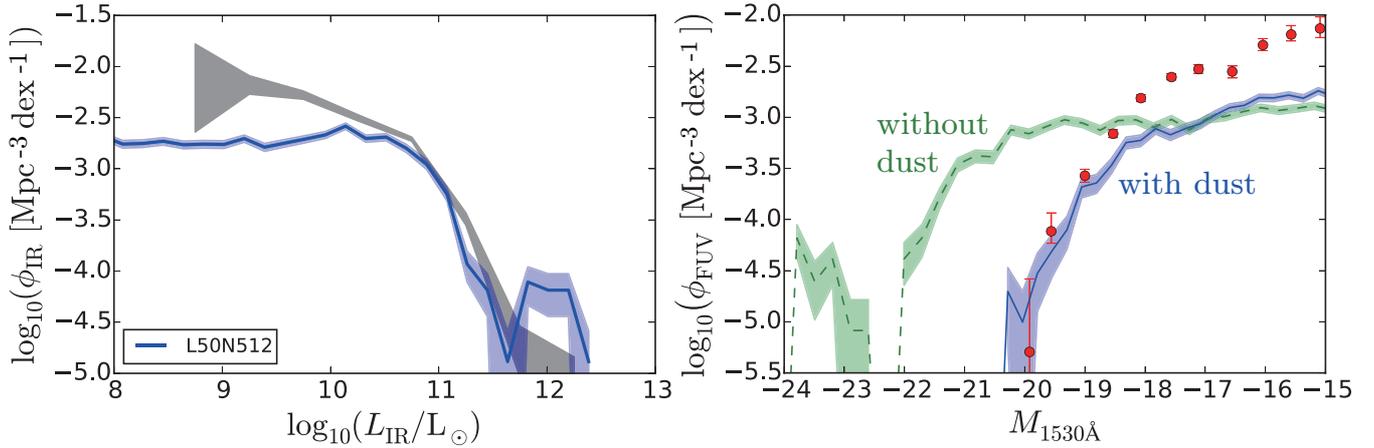}
        \caption{{ {\it Left panel:}  IR LF  in the simulation at $z=0$ with the shaded
        region showing the Poisson error. The grey shaded region indicates the IR LF obtained by {\it Herschel} with 1 $\sigma$ uncertainty \citep[][]{2013MNRAS.432...23G}.
        {\it Right panel:} Rest-frame UV LF in the simulation at $z=0$.
        The solid and dashed lines show the results with and without dust extinction.
          The blue and green shaded regions show the Poisson error.
 Observational result from {\it GALEX} is shown as red points with error bars \citep[][]{2005ApJ...619L..19T}.}
    }
    \label{fig:luminosity_function}
\end{figure*}

\subsection{Radius of IR-emitting region}\label{subsec:Rdust-LIR}

In our simulation, the radius of IR-emitting region is assumed to be
$\alpha$ times the scale length of dust distribution ($R_\mathrm{dust}$).
We determine $\alpha$ to reproduce the observed IR-emitting region sizes
of IR-luminous galaxies with $L_\mathrm{IR}\gtrsim 10^{12}$ L$_{\sun}$.
We find that $\alpha =2$ can reproduce the ALMA measurement
of IR-emitting region sizes in \citet[][]{2017ApJ...850...83F} as shown in Fig.~\ref{fig:R_IR0}; 
we compare the distribution of $\alpha R_\mathrm{dust}$ with $\alpha =2$
for the simulated galaxies whose IR luminosity is greater than
$10^{11}\,{\rm L}_{\odot}$ with the observational data at $z=1$--4.
We find that our dust-emitting region radii agree not only with the observations
at a particular redshift, but also with the observed decreasing trend 
towards higher redshifts.
This comparison shows that our simulation captures 
the redshift evolution of dust-emitting region consistently with the current observation. 
Because $\alpha R_\mathrm{dust}$ is important in determining
the dust optical depth (or the dust surface density; equations~\ref{eq:surface_dust}
and \ref{eq:tau}),
the consistency in $R_\mathrm{dust}$ confirms that our simulation can give
a reasonable estimate for the dust optical depth under a given dust mass.

\subsection{IR LF}\label{subsec:LF_z0}

The IR LF is significantly affected by the dust abundance and
the stellar populations of galaxies. 
Thus, comparison of the IR LF in the simulation with the observed one 
is useful to test our treatment of dust evolution and star formation as shown in 
Fig.~\ref{fig:luminosity_function}, where we compare our simulation results with the observational data
obtained by {\it Herschel} at $z=0$ \citep[][]{2013MNRAS.432...23G}.
The IR LF is also derived from other IR observations
(e.g.\ \citealt{2018MNRAS.474.5363K} for a recent `accurate' IR LF).
Since recently obtained IR LFs at $z\sim 0$ are consistent with the previous results within the uncertainty of our theoretical estimates of $L_\mathrm{IR}$,
we only focus on the \textit{Herschel} data in this paper.

{ We also compared the UV LF at $z\simeq 0$ with a corresponding observational result \citep[{\it GALEX}{\rm ;}][]{2005ApJ...619L..19T} in Fig.~\ref{fig:luminosity_function}. We observe a good agreement at the bright-end of $M_{1530}<-19$, where $M_{1530}$ is the rest-frame UV magnitude at 1530\,\AA. However we underestimate the number of UV faint objects at $M_{1530}>-19$ by 0.5 dex at most.  
This is because the low-mass galaxies in our simulation is not so active in star formation.
We also show the intrinsic UV LF (before applying dust extinction) in Fig.~\ref{fig:luminosity_function} in order to show the impact of dust.  If dust extinction effect is not considered, then the bright end of UV LF is overestimated and the observed data points cannot be explained.
We emphasize that the dust extinction correctly suppresses the UV luminosity at the
bright end, which is important in our studies of IR-luminous galaxies.}

{ There is a bump in the IR LF at $L_{\rm IR}\simeq 10^{12}\,L_{\odot}$, which corresponds to the bump at $M_{1530}\simeq -23.5$ in the intrinsic UV LF. 
This bump is created by 
just a few objects in the simulation box.  In this simulation, the star formation in some massive galaxies is not quenched even in the low-$z$ Universe. However, this bump does not affect the conclusion of this paper because the number of such massive galaxies is very small.}

Our simulation reproduces the observed IR LF within a factor of 3
at $z=0$, although the simulated LF is lower than the observed data at the faint end. 
The overall match of LF indicates that our model is
`calibrated' roughly by the local observation in a statistical sense. 
In other words, the total IR emissivity in a cosmological volume
is correctly modeled in our simulation; i.e., the total fraction of
stellar light absorbed by dust is appropriately modeled.
This success at $z=0$ gives a basis on which we discuss the LFs at other redshifts.

{ We also investigated the relation between $L_{\rm IR}$ and $\mathcal{D}_{\rm tot}$. $\mathcal{D}_{\rm tot}$ and $L_{\rm IR}(z)$ are correlated at all redshifts ($z < 5$). $L_{\rm IR}(z)=10^{12}\, L_{\odot}$ corresponds to $\mathcal{D}_{\rm tot} = 0.01$. When one fixes $\mathcal{D}_{\rm tot}$, $L_{\rm IR}(z)$ decreases as the redshift increases. The reason is that the stellar population becomes older as the redshift becomes smaller.
}

However, the LF only provides a statistical aspect of the dust emission
properties.  In the following subsections, we examine the relations among quantities
characteristic of dust emission in individual galaxies at $z=0$.

\subsection{$T_{\rm dust}-L_{\rm IR}$ relation}
\begin{figure}
	\includegraphics[width=\columnwidth]{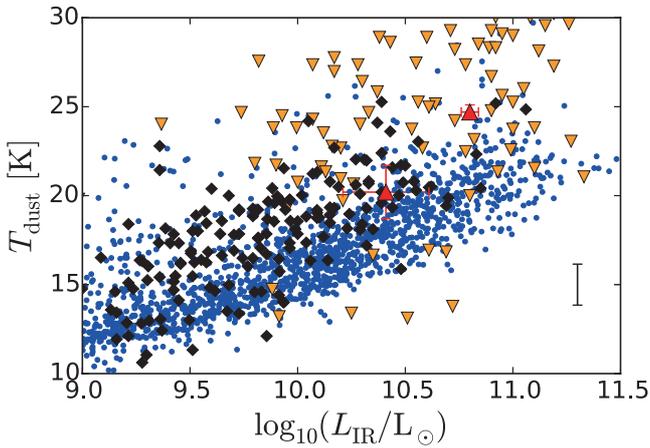}
    \caption{
      Relation between dust temperature and IR luminosity at $z=0$.
      The simulated galaxies are represented by the blue points. The
      black diamonds, 
      orange inverted triangles 
      and red triangles 
      are taken from \citet{2013MNRAS.433..695C}, \citet{2010A&A...518L...9A} and
      \citet{2018MNRAS.475.5585Z}, respectively.
      The typical observational uncertainty of the dust temperature is shown in the bottom
       right corner \citep{2013MNRAS.433..695C}.
}
    \label{fig:T_IR0}
\end{figure}

Dust temperature depends on the total radiative energy incident on the dust.
Therefore, the $T_{\rm dust}-L_{\rm IR}$ relation gives us a useful clue as to
how the increase of dust emission is associated with the increase of
dust heating. In Fig.~\ref{fig:T_IR0},
we compare the $T_{\rm dust}-L_{\rm IR}$ relation in the simulation
with the observational data for local galaxies \citep[][]{2013MNRAS.433..695C, 2018MNRAS.475.5585Z}.
We successfully and quantitatively reproduce the observed trend of rising
$T_\mathrm{dust}$ with increasing $L_\mathrm{IR}$ seen for most of the sample.
This means that the stellar radiation field incident on the dust (or the spatial
distribution of stars and dust) is appropriately modeled in the simulation.

\subsection{IRX--$\beta_{\rm UV}$ relation}\label{subsec:IRX-beta_z0}

Finally, we examine the IRX--$\beta_\mathrm{UV}$ relation, which connects the
dust extinction and emission properties. 
As shown in Fig.~\ref{fig:IRX-beta0}, 
our simulation shows a rising trend of IRX and $\beta_\mathrm{UV}$,
which roughly agrees with the observed trends represented by
\cite{1999ApJ...521...64M}, \cite{2012ApJ...755..144T} and \cite{2014ApJ...796...95C}
(see also references therein).
A significant fraction of galaxies are distributed around the curves which are obtained
for a nearby galaxy sample.
A large part of objects with low stellar mass
($10^{8}  \lesssim M_{\ast}/{\rm M}_{\odot} \lesssim 10^{9} $; dark blue points in
Fig.\ \ref{fig:IRX-beta0})
have low IRX and relatively red SEDs
because the 
dust-to-gas mass ratio of these galaxies is small ($\mathcal{D}_{\rm tot}\lesssim 10^{-3}$)
(i.e.\ the dust extinction is small)  and the current star formation activity is low
(i.e.\ the stellar colour is red) in our simulation.

There are several very actively star-forming galaxies seen in their
extremely blue colours around $\beta_{\rm UV}\simeq -3$.
They are dust-poor low-mass systems whose stellar light is dominated by
a very young ($\sim 10^6$~yr) stellar population and is little extinguished by dust.
Massive starburst galaxies show high IRX ($\gtrsim 10$) and $\beta_\mathrm{UV}$ between
$\sim -2$ and $-1$.
The narrow range of $\beta_\mathrm{UV}$ is due to  the limit of $\beta_{\rm UV}$
in the mixed dust geometry as we discuss
in Section~\ref{subsec:geometry} (and Appendix \ref{app:beta_limit}).
Thus, the range of $\beta_\mathrm{UV}$ for those massive starbursts may be
underestimated. If we assume a screen geometry, the data points are distributed
over a wider range of $\beta_\mathrm{UV}$, but the IRX of some galaxies
also becomes extremely high
(up to $10^6$; such a high number is not observed).
Probably, the real geometry of dust distribution is in the middle of
these two extremes (mixed and screen). Since we do not resolve the detailed dust
distribution, we do not tune the dust geometry. We only emphasize that we
obtain the IRX--$\beta_\mathrm{UV}$ relation broadly consistent with the observed
one.

\section{Comparison with the STUDIES Sample}\label{sec:STUDIES}

\subsection{Observational data}\label{subsec:sample}

We compare our simulation results with the high-redshift observational sample
obtained by the STUDIES project. We adopt deep submm imaging surveys that have been
carried out in the COSMOS field with SCUBA-2 by various teams
\citep[][]{2013MNRAS.436.1919C, 2013MNRAS.432...53G, 2017MNRAS.465.1789G, 2017ApJ...850...37W}
at 450 and 850 $\micron$. \cite{2013MNRAS.436.1919C} performed
a large and uniform blank-field survey over a large scan area of $\sim 15'\times 15'$
with a noise level of $\sim 3.6$ mJy {at 450 $\micron$}.
The SCUBA-2 Cosmology Legacy Survey \citep[S2CLS;][]{2013MNRAS.432...53G, 2017MNRAS.465.1789G}
and the SCUBA-2 Survey of SMGs in the COSMOS field
(S2COSMOS; Simpson et al. in preparation)
covered several well studied extragalactic legacy fields. We adopt the deep
blank-field survey in the COSMOS field. The noise level at the deepest regions
of this survey is $\sim 0.9$ mJy at 450 $\micron$.  Finally, STUDIES
\citep{2017ApJ...850...37W}
is an ongoing JCMT Large Program that aims at reaching the confusion
limit in the COSMOS-CANDELS field.
The current r.m.s noise at the deepest regions of this image is $\sim 0.75$ mJy. 
After combining all these data, the extremely deep image leads to a sample of 269 galaxies selected at 
  450 $\micron$ \citep[see, e.g.,][]{2018arXiv180807480C}, with a redshift range of $0.21< z <4.37$ (median $z=1.64$).
{In addition to the dataset of STUDIES, we also use the observational data
obtained by another SCUBA-2 project with a comparable capability \citep{2018MNRAS.475.5585Z}
and a representative {\it Herschel} program \citep[][]{2014ApJ...796...95C} for reference.
Note that these two samples give consistent conclusions to those derived from
the STUDIES sample as we see below.}

Since the COSMOS field has been surveyed at other wavelengths,
Lim et al. (in preparation; see also \citealt{2018arXiv180807480C}) used those data to derive
physical properties
of the sources with SED fitting in the optical and IR. As a
consequence, they succeeded in deriving the relations between the quantities 
that we considered in the previous section such as
IRX, $\beta_{\rm UV}$, dust temperature, and IR luminosity.
The sample provides us with
an opportunity of comparing our theoretical results with a uniform sample
(sample collected by a uniform method), so that we do not have to worry about
systematic difference between different samples.
We should keep in mind that the sample we adopt is IR-selected:
Those submm sources are usually optically faint mainly because
they are heavily obscured by dust.

\begin{figure}
	\includegraphics[width=\columnwidth]{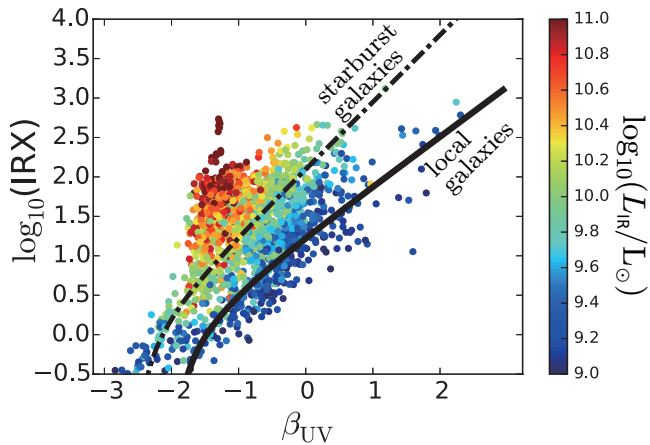}
    \caption{
    IRX--$\beta_{\rm UV}$ relation at $z=0$.
    The colour of each point shows the 
    logarithmic IR luminosity in units of ${\rm L}_{\odot}$ 
    as shown by the colour bar.
    We show the fitting formulae of IRX--$\beta_{\rm UV}$ for starburst galaxies
\citep[dot--dashed line;][]{2004MNRAS.349..769K} and
galaxies in the local Universe \citep[solid line;][]{2012ApJ...755..144T}. 
}
    \label{fig:IRX-beta0}
\end{figure}
Below we compare the IR-emitting properties of the simulated galaxies at various
redshifts to the above observational sample.
We consider the redshift ranges $0.5<z<1.5$, $1.5<z<2.5$, $2.5<z<3.5$, and
$3.5<z<4.5$, where the observed LFs are derived with sufficient statistical significance. 
We adopt the simulation snapshots at $z=1$, 2, 3, and 4 for comparison. 

\subsection{$M_*$--$L_\mathrm{IR}$ relation}

\begin{figure}
	\includegraphics[width=\columnwidth]{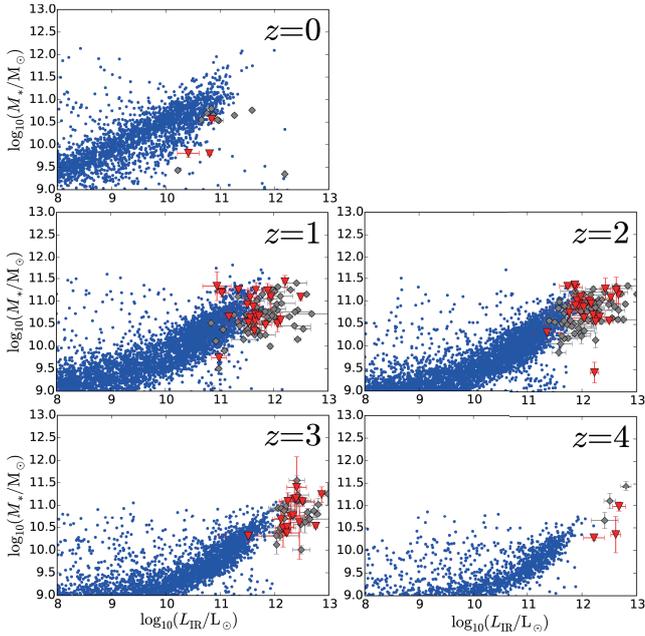}
    \caption{
    Relation between stellar mass and IR luminosity in the
    simulated galaxies (blue points) and the observational sample
    [gray diamonds ($\diamondsuit $; Lim et al., in preparation) and red triangles
    ($\triangledown $; \citealt{2018MNRAS.475.5585Z})] at $z=0$, 1, 2, 3, and 4.} 
    \label{fig:LM}
\end{figure}

Before examining the dust emission properties, it is useful to clarify
the stellar mass ($M_*$) in the IR luminosity range covered by the STUDIES sample.
In Fig.~\ref{fig:LM}, we show the relation between IR luminosity
and stellar mass of each galaxy in the simulation.
Although the scatters are large, there is a clear correlation between
these two quantities. 
We also show the observational points which are obtained by STUDIES (Lim et al. in prep.) and SCUBA-2 data \citep[][]{2018MNRAS.475.5585Z}
as gray diamonds and red triangles with error bars, respectively.

The simulated galaxies are consistent with the observed data at the massive end, 
and we see that our simulation contains numerous lower mass galaxies that are 
currently not observed by the IR surveys (i.e.\ below their observational limit). 
At $z>3$, our simulation lacks the very IR-luminous objects 
with $L_{\rm IR}>10^{12}\,{\rm L}_{\odot}$ as we also discuss in Section \ref{subsec:LF_z}.

\subsection{IR luminosity function}\label{subsec:LF_z}
\begin{figure}
	\includegraphics[width=\columnwidth]{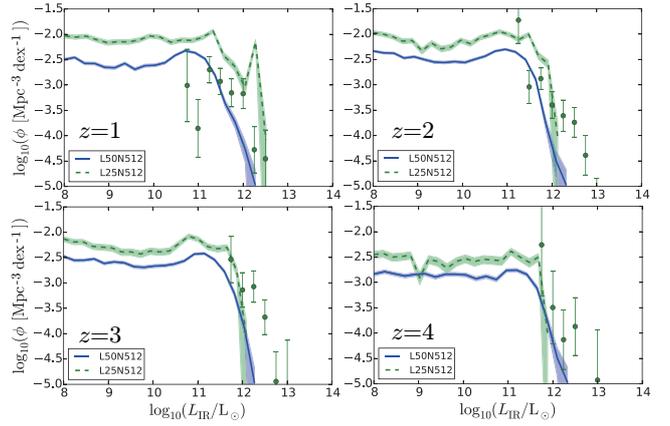}
    \caption{
    IR luminosity functions at $1 \le z \le 4$.
    The solid blue and dashed green lines represent the results of L50N512 and L25N512, respectively.
    The shaded regions show the Poissonian error. 
    The points with error bars are the STUDIES data. 
    }
    \label{fig:LF_z}
\end{figure}

We show the IR LFs at $z=1$--4 in Fig.~\ref{fig:LF_z}.
The LFs in the default run (L50N512) are consistent with the data
at their `knees', while they tend to underproduce the abundance of galaxies
at $L_\mathrm{IR}>10^{12}$ L$_{\sun}$.
Thus, our default simulation is not fully successful in treating
a mechanism of producing extremely IR-luminous objects.

There are some possible reasons for the failure at the IR-luminous
end.
The lack of extremely IR-luminous objects could be due to
insufficient spatial resolution (see Section~\ref{subsec:res}).
To present the effect of spatial resolution,
we show the results for the two box sizes (L50N512 and L25N512).
Indeed, the higher-resolution run is successful in producing galaxies with
$L_\mathrm{IR}>10^{12}$~L$_{\sun}$ at $z=1$ because more
concentrated (or denser) starbursts are realized in the high-resolution condition.
The higher-resolution run reproduces the observed luminosity functions up to
$L_\mathrm{IR}\sim 10^{12}$ L$_{\sun}$,
but still fails to explain the brightest ($L_\mathrm{IR}\gtrsim 10^{13}$ L$_{\sun}$) objects
detected at $z>2$ (despite the better physical-scale resolution at higher redshift).
At $z\gtrsim 2$, the major effect of increased spatial resolution is to increase 
the abundance of galaxies below the knee of luminosity function, i.e., the lower mass galaxies in lower mass halos. 
Therefore, we conclude that the poor spatial resolution is not the only reason
for not reproducing the extremely IR-luminous objects at $z\gtrsim 2$.

It is worth emphasizing the good match between the simulation and observation
at the knees of luminosity functions. This is achieved by the depth of the STUDIES sample.
However, we should keep in mind that
it is still difficult to theoretically reproduce the brightest objects with $L_\mathrm{IR}\gtrsim 10^{13}$ L$_{\sun}$.
We discuss a possibility of explaining such luminous objects with AGN contribution in Section~\ref{subsec:AGN}.

\subsection{$T_{\rm dust}-L_{\rm IR}$ relation}

\begin{figure}
	\includegraphics[width=\columnwidth]{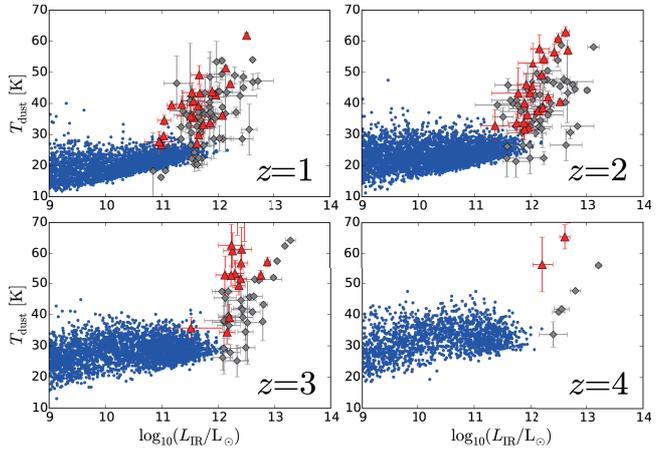}
        \caption{
    $T_{\rm dust}$-$L_{\rm IR}$ relation at $z=1$--4.
          The grey diamonds 
          and red triangles 
          with error bars are taken from the STUDIES sample (Lim et al., in preparation) and
          another SCUBA2 sample \citep[][]{2018MNRAS.475.5585Z}, respectively.
    }
    \label{fig:T_Lfir_z}
\end{figure}

In Fig.~\ref{fig:T_Lfir_z}, we plot the $T_{\rm dust}$--$L_{\rm IR}$
relation at $z=1$--$4$ with observational results from
STUDIES (Lim et al. in prep.)
and {\it Herschel} \citep[][]{2014ApJ...796...95C}.
Because the observations only detect luminous objects
$(L_{\rm IR}\gtrsim 10^{11}L_{\odot}~{\rm at}~z=1$ and
$L_{\rm IR}\gtrsim 10^{12}L_{\odot}~{\rm at}~z=3)$,  they can only be
compared with the luminous population in our simulation.
The dust temperatures of the luminous galaxies in our simulation 
are within the dispersion of the observed ones at all redshifts.
The observational data seem to have greater dispersion than the
theoretical predictions.  The diversity in the small-scale dust
distribution, which cannot be resolved in our simulation, could cause 
more dispersion in the dust temperature.  Since a region with a higher
dust temperature emit more, the inhomogeneity tends to raise the apparent
$T_\mathrm{dust}$ \citep[e.g.][]{2011MNRAS.411.1707S}
which could explain high-$T_\mathrm{dust}$ objects in the observational sample.
We also find that the dust temperatures in the simulation tend to become higher as the redshift increases, which is not clearly seen in the observational sample that we are comparing against. 
However, we note that some observational studies showed that the dust temperatures
tend to be higher at higher redshifts \cite[][]{2018A&A...609A..30S}.

In the simulation, we do not have sufficient number of galaxies with
$L_\mathrm{IR}\gtrsim 10^{13}$ L$_{\sun}$, but the extrapolation of
the luminous end successfully explains the observed dust temperature.
As we will discuss in Section~\ref{subsec:AGN}, there could be an extra source of dust heating.
Since the simulation box contains galaxies up to $L_\mathrm{IR}\sim 10^{12.5}$ L$_{\sun}$,
the increase of $L_\mathrm{IR}$ by a factor  of 3 would be necessary.
If we raise $L_\mathrm{IR}$ by a factor of 3, $T_\mathrm{dust}$ would be
$3^{1/6}\simeq 1.2$ times higher (Eq.~\ref{eq:IR4}). Because the
change in the dust temperature is small, the
good match of $T_\mathrm{dust}$ between the simulation and observations
is not significantly affected by the underestimate of $L_\mathrm{IR}$.
Therefore, we conclude that the dust temperatures obtained in the simulation
lie in a range consistent with the observed values.

\subsection{IRX--$\beta_{\rm UV}$ relation}

\begin{figure}
  \includegraphics[width=\columnwidth]{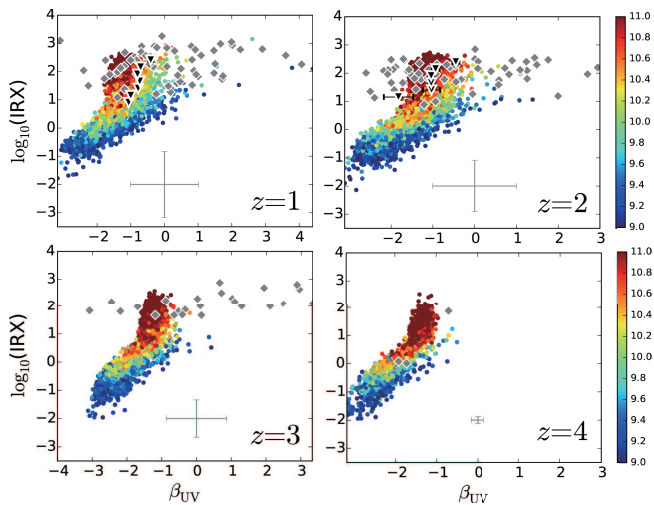}
  \caption{
    IRX--$\beta_{\rm UV}$ relation at $1\le z \le 4$. The colour bar
    indicates the logarithmic IR luminosity in units of ${\rm L}_{\odot}$ of each galaxy.
    The grey diamonds 
    and black triangles 
    show the observational samples (STUDIES data taken from Lim et al., in preparation
    and {\it Herschel} data from
    \citealt{2014ApJ...796...95C}).
    Note that \citet{2014ApJ...796...95C} took the average of $\beta_{\rm UV}$
    for each IRX bin.  The typical size of error bars for the STUDIES sample is shown in each panel.}
    \label{fig:IRX-beta_z}
\end{figure}

We show the IRX--$\beta_{\rm UV}$ relations at $z=1$--4 in Fig.~\ref{fig:IRX-beta_z}.
Because the IR-selected STUDIES sample is optically faint, the determination of $\beta_{\rm UV}$ is uncertain (i.e.\ error bars are large).
Our simulation reproduces the observed IRX--$\beta_\mathrm{UV}$ relation in the
following two aspects. First,
the IRX values are broadly consistent with the observed ones.
Since the observational sample is IR-selected, it only covers the high IRX values.
Second, the $\beta_\mathrm{UV}$ values predicted by our simulation are roughly located in the middle of the observed range
of $\beta_\mathrm{UV}$ for $z=1$--3, although the uncertainty in the observed
$\beta_\mathrm{UV}$ is large. The sample size is too small at $z=4$.

Although it is difficult to derive an evolutionary trend from the
scattered observational data, the theoretical results predict some evolutionary tendencies.
As the redshift becomes higher, the extension of the simulation data toward large
$\beta_\mathrm{UV}$ is less mainly because of less contribution from
old stellar populations.
The clump of data located between $\beta_\mathrm{UV}\sim -2$ and $-1$ is 
also seen at $z=0$ (Fig.\ \ref{fig:IRX-beta0}), and is
due to the optically thick limit of $\beta_{\rm UV}$ (see Appendix \ref{app:beta_limit}).
Although the observational error bars are large, there is a tension between theoretical
and observational $\beta_\mathrm{UV}$ values in that the simulation underproduces
galaxies with $\beta_\mathrm{UV}>0$. This may be partly resolved by changing the dust
distribution geometry. As mentioned in Section \ref{subsec:IRX-beta_z0},
if we assume a screen geometry, $\beta_\mathrm{UV}$
can become larger; however, we should keep in mind that IRX also becomes larger
(note that the observed IRX values are nicely reproduced by the simulation).
Because of the large uncertainties in the observational
$\beta_\mathrm{UV}$ values,
we do not fine-tune the simulation results any further.

\section{Discussion}\label{sec:discussion}

Although our simulation is successful in explaining some important aspects
in the dust emission properties of the STUDIES sample, there are some
discrepancies and uncertainties. In this section, we give further discussions on these issues. 

\subsection{Effect of spatial resolution}\label{subsec:res}

It is suggested that the brightness of a starburst is limited to
$\lesssim 10^{13}$ L$_{\odot}$ kpc$^{-2}$
by radiation pressure (or the Eddington limit)
\citep{2018arXiv180203117C}.
Ideally, the simulation should be able to treat such a high-density starburst.
However, in our simulation (or in cosmological simulations under the current computational
resources), it is practically impossible to realize such an extreme starburst, as we estimate
in what follows.

The highest gas density that can be achieved in the simulation is related to
the smallest smoothing length of gas particles, $h_{\rm min}$. 
Since $h_{\rm min}\simeq 0.3$ kpc (physical) at $z=3$ 
for L50N512, only about 40 gas particles can be packed in a $\sim$1\,kpc region.
When we consider the situation that these gas particles with a mass of 
$7.0\times 10^{8}\,{\rm M}_{\odot}$  are converted into stars within a free-fall time 
of $\simeq 10^{7}$\,yr, 
the intrinsic UV luminosity is estimated to be $2.5\times 10^{11}\,{\rm L}_{\odot}$  
based on the population synthesis model we adopted.
Therefore, if we consider that all UV photons are reprocessed into the IR by dust grains,
the IR luminosity roughly corresponds to the cut-off scale of luminosity function in
Fig.\,\ref{fig:LF_z}. In this sense, the lack of extremely IR-luminous objects could be
due to the limited spatial resolution.

Indeed, as discussed in Section \ref{subsec:LF_z},
a higher spatial resolution serves
to produce luminous IR galaxies at $z=1$. 
However, a higher spatial resolution requires a smaller box size 
under a fixed computational resource. Since IR-luminous objects are rare,
a large box size is also important. For example, if we halve the comoving box size to
$12.5\,h^{-1}$\,Mpc from  $25\,h^{-1}$\,Mpc, 
the objects with comoving number densities  $\lesssim 10^{-3.5}$ Mpc$^{-3}$ cannot be
produced, while the observations detect rare objects with $\sim 10^{-5}$\,Mpc$^{-3}$.
{  On the other hand, increasing the box size with a fixed particle number would reduce the resolution, and then the central region of IR-luminous objects such as compact starburst or AGN cannot be resolved. 
Ideally one would like to increase the box size and resolution at the same time with a larger number of particles; however, this requires far more computing resources, and it is not feasible at this moment. 
}

Zoom-in simulations, which utilise higher resolution gas particles selectively in the region of interest,
are a viable way of achieving a high resolution in cosmological simulations
\citep[e.g.][]{2015MNRAS.451..418Y,2018MNRAS.480..800H}.
However, since the regions of interest are pre-selected, this method is not suitable for predicting
statistical quantities such as luminosity function. On the other hand, it may be possible to
apply a zoom-in method to one of the IR-brightest objects. This will enable us
to test whether a dense gas concentration actually leads to an extremely IR-luminous objects or not.

\subsection{Geometry of dust distribution}\label{subsec:geometry}

In the above, we assumed a mixed distribution of dust and stars.
However, because there is an upper limit for $\beta_\mathrm{UV}$
as shown in Appendix \ref{app:beta_limit}, $\beta_\mathrm{UV}$ could
be underestimated in the mixed geometry. We could alternatively
consider screen geometry, which does not have any upper limit for $\beta_\mathrm{UV}$. 
\cite{2002ApJ...570..470T} considered in their `screen model' that
only a part of dust is heated by stellar radiation. They explained
the $T_\mathrm{dust}$--$L_\mathrm{IR}$ relation by the screen geometry rather than the `slab' geometry. 
Our mixed geometry is close to their slab treatment.
However, in our model, if we adopt the screen geometry, 
we find that some galaxies reach as high as IRX $\sim 10^{6}$. 
Since no observed galaxies show such a large value, 
the screen geometry is not suited for our purpose.

As mentioned above, a zoom-in simulation focusing on IR-luminous galaxies could
resolve this issue, since we do not need to assume a specific relative distribution
between dust and stars under a high enough spatial resolution.
The highest zoom-in simulation today reach resolution of a few pc 
\citep[e.g. FIRE-2 simulations by][]{2018MNRAS.480..800H}, 
so the relative geometry of dust and gas can be resolved down to 
such scales if a dust model is implemented. 
Nevertheless, there could be even finer clumpy structure below a few pc, 
which warrants future research to clarify how such small-scale structures could play 
a role in the absorption and reemission by dust.

\subsection{Variation of extinction curves}\label{subsec:var_ext}
\begin{figure}
	\includegraphics[width=\columnwidth]{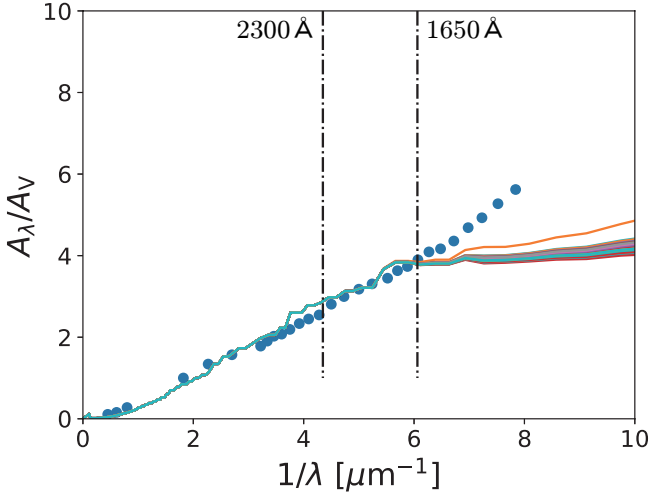}
    \caption{Extinction curves normalized to $A_V$ for galaxies with
    $L_\mathrm{IR}> 10^{11}\,{\rm L}_{\odot}$ at $z=0$ (solid lines). 
      The blue circles indicate the SMC extinction curve taken from
      \citet{1992ApJ...395..130P}.
      We mark the rest-frame wavelengths
      used to determine $\beta_\mathrm{UV}$ by the vertical dot-dashed lines
      (2300 {\AA} on the left and 1650 {\AA} on the right).}
    \label{fig:Ex}
\end{figure}

One of the new features of our dust treatment is that we included the grain size distribution in the form
of the two-size approximation. Therefore, the extinction curve is a predicted quantity
in our model. We show the extinction curves of IR-luminous
galaxies in the simulation in Fig.\ \ref{fig:Ex}. 
We only focus on galaxies with $L_\mathrm{IR}>10^{11}\,{\rm L}_{\odot}$ { at $z=0$},
which roughly corresponds to the luminosity range of the observational sample that we are comparing to.
 We normalize all extinction curves to $A_V$
(the extinction at $\lambda =0.55~\micron$) in order to focus on their shape.
We also plot the SMC extinction curve taken from
\cite{1992ApJ...395..130P} in Fig.~\ref{fig:Ex}.
The extinction curves in the simulation are consistent with the SMC curve at $\lambda > 1650$\,\AA.
At shorter wavelengths,  the extinction is suppressed, which means
that the grain abundance is dominated by large grains in those
galaxies.
This is due to efficient coagulation in the central parts of massive galaxies
\citep[see][]{2017MNRAS.466..105A, 2017MNRAS.469..870H}.
In our model, the value of $\beta_\mathrm{UV}$ is determined in the wavelength range
where the extinction curve is approximated well by the SMC curve.
{ In our separate work, Hou et al.\ (submitted) discuss the evolution of extinction curves. They showed that, if we focus on solar-metallicity ($Z\gtrsim 0.3$ Z$_\odot$)
objects, which are valid for
the IR-luminous objects in our simulation, the extinction curve does not evolve significantly
over the redshift range.
We confirmed that the extinction curves of IR-luminous objects
do not change significantly along the redshift also in our simulation.
}

\subsection{More dust heating by AGN and massive stars at high $z$?}\label{subsec:AGN}

\begin{figure}
	\includegraphics[width=\columnwidth]{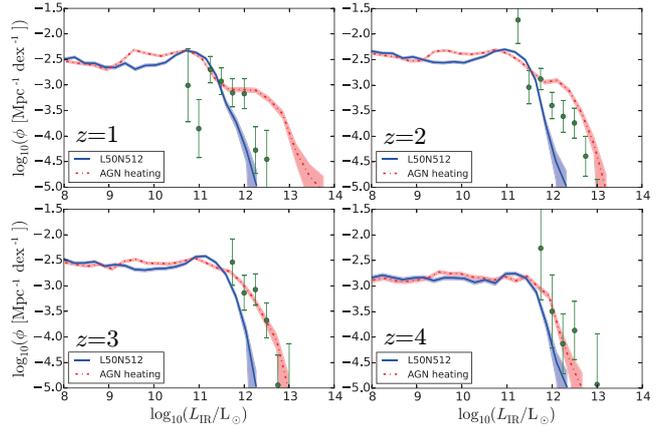}
        \caption{
          Possible impact of the extra dust heating by AGNs on the IR luminosity function at $z=1$--4.
          The red dotted and blue solid lines show the IR luminosity function with and without 
          the AGN effect, respectively. The shaded region represent the 1-$\sigma$ uncertainty 
          in Poisson statistics. The data points are the same as in Fig.~\ref{fig:LF_z}.
    }
    \label{fig:AGN}
\end{figure}

In this paper, we only considered dust heating by stellar radiation.
If AGNs contribute significantly to dust heating, the underproduction of
the IR luminosity function could be explained by the lack of AGN treatment in our simulation.
\cite{2013MNRAS.432...23G} estimated the fraction of AGN for galaxies ($0 \le z \le 4$)
detected by {\it Herschel}.
After performing spectral analysis in a wide wavelength range
($0.1~\micron\le \lambda\le 1000~\micron$), they concluded that approximately 50 per cent of observed galaxies are classified as galaxies hosting an AGN.
Chiang et al. (submitted) have shown that the number fraction of galaxies hosting an AGN
is roughly 60 per cent for $L_\mathrm{IR}>10^{12}$~L$_{\sun}$ and that the fraction does not
change along redshift up to 2.
On the other hand,  the database of \cite{2017ApJS..233...19C} showed that approximately 10 per cent of galaxies have AGNs at the redshift range $2.0 \leq z \leq 2.5$.
Moreover, the contribution from AGNs is usually prominent at rest-frame mid-IR wavelengths
according to the SED fitting by \citet{2017ApJS..233...19C}; thus, it is not obvious
if the above percentage reflects the fraction of AGNs contributing to the STUDIES bands.
Recently, \cite{2018ApJ...857...31T} discovered an extremely IR-luminous
galaxy ($L_\mathrm{IR}\sim 2\times 10^{14}$ L$_{\sun}$) at $z=3.7$, 
in which a significant fraction of far-IR emission is contributed by an AGN, although the star formation  has a comparable contribution to the AGN at wavelengths relevant for the STUDIES observations.

Although robust quantification of AGN contribution is still a matter of debate, we
attempt the following extreme corrections in order to investigate the possible
maximum contribution from AGNs to the total IR luminosity.
We assume that a fraction super massive blackholes (SMBHs) manifest AGN activity
and that the luminosity of the AGN is described by the Eddington luminosity of the SMBH.
For simplicity, we also assume that all the radiation from the AGN is 
reprocessed into IR photons by surrounding dust grains.

The SMBH mass ($M_{\rm BH}$) can be estimated by the so-called Magorrian relation \citep[][]{1998AJ....115.2285M}, and we calculate it according to \cite{2004ApJ...604L..89H} as 
\begin{eqnarray}
M_{\rm BH}&=&10^{8.2}\left(\dfrac{M_{\rm bulge}}{10^{11}\,{\rm M}_{\odot}} \right)^{1.12}\,{\rm M}_{\odot}\,,
\end{eqnarray}
where $M_{\rm bulge}$ is the bulge mass of the galaxy.
Since our simulation is not capable of resolving the galaxy morphology in detail,
we simply use the total stellar mass for $M_\mathrm{bulge}$.
The Eddington luminosity, $L_\mathrm{Edd}$ is estimated by \citep[e.g.][]{1979rpa..book.....R}
\begin{eqnarray}
  L_\mathrm{Edd} &=&\dfrac{4\pi GM_{\rm BH} m_{\rm p}c}{\sigma_{\rm T}}=5.07 \times 10^{12}
  \left(\dfrac{M_{\rm bulge}}{10^{11}\,{\rm M}_{\odot}} \right)^{1.12}\,{\rm L}_{\odot}\,,\notag \\
\end{eqnarray}
where $G$, $m_{\rm p}$ and $\sigma_{\rm T}$ are
the gravitational constant, the mass of proton, and the Thomson scattering cross-section of electron, respectively.
We add $L_\mathrm{Edd}$ to the IR luminosity which originates from
the dust reprocessing of stellar light (equation~\ref{eq:reprocessing}) and regard the sum as the
total IR luminosity for the galaxies which are chosen as AGN hosts.

We select galaxies hosting an AGN randomly with a probability given by the
AGN fraction.
The AGN fraction $f_{\rm AGN}$ is observationally investigated by
Chiang et al.\ (submitted) in a wide redshift range of $0 \le z \le 2.5$.  
They show that $f_{\rm AGN}$ is highly dependent on the IR luminosity
and is almost independent of the redshift: 
$f_{\rm AGN}\simeq 0.6$ for $L_{\rm IR} \geq 10^{12}\,{\rm L}_{\odot}$,
$f_{\rm AGN}\simeq 0.4$ for $10^{11}\,{\rm L}_{\odot} \leq L_{\rm IR} < 10^{12}\,{\rm L}_{\odot}$ and
$f_{\rm AGN}\simeq 0.1$ for $L_{\rm IR} < 10^{11}\,{\rm M}_{\odot}$.
We adopt the these $L_\mathrm{IR}$-dependent fractions, assuming that
$f_\mathrm{AGN}$ does not depend on the redshift. 

In Fig.\ \ref{fig:AGN}, we show the IR luminosity functions with and without the AGN contribution. 
The extremely IR-luminous galaxies ($L_{\rm IR}\sim 10^{12.5}\,{\rm L}_{\odot}$)
are well accounted for by the additional AGN contribution at $z>1$ in the L50N512 run. 
However,  we overestimate the abundance of IR-luminous galaxies 
at $z=1$.   Since A18's simulation 
does not include AGN feedback, we probably need to include a consistent
treatment between AGN feedback and emission in the future. 
Note again that our inclusion of AGN emission is based on an `extreme' assumption
that all the Eddington luminosity is reprocessed in the IR and that $M_\mathrm{bulge}$
is equal to the total stellar mass.

Another possible `extra' heating could be due to a top-heavy IMF,
which boosts the total stellar luminosity especially in the UV.
\cite{2005MNRAS.356.1191B} argued that their semi-analytic model, which is consistent
with the
Lyman-break galaxy luminosity function and the properties of the local galaxy population in the optical and IR, can explain the observed galaxy number counts at 850 $\mu$m 
only in the case of a top-heavy IMF.
However, \cite{2013MNRAS.428.2529H} used an empirical model based on simulations including galaxy--galaxy interactions, and suggested that the observed submm galaxy number counts do not provide evidence for a top-heavy IMF at high redshift.
\cite{2018Natur.558..260Z}
have recently shown that the isotopic ratios of CO in massive starbursts
support a top-heavy IMF.
Because there is a large freedom in modeling a top-heavy IMF, 
we only mention that a top-heavy IMF is a possible way of
reproducing the extremely IR-luminous objects.

\section{Summary and Conclusions}\label{sec:conclusion}

We examine the consistency (and inconsistency) between 
the cosmological simulation with dust evolution  
and the deepest IR-selected observational sample.
We choose the STUDIES sample for the observational data
because of its depth and coverage at multiple wavelengths.
We use the cosmological hydrodynamical simulation by A18,
in which we computed the dust evolution consistently with the star formation activities and
the local physical conditions of the ISM. We adopted the
two-size approximation of grain size distribution \citep[][]{2015MNRAS.447.2937H},
in which the entire grain radius range is represented by two sizes.
In the simulation, we are able to trace spatial structures down to $\sim 1$ comoving kpc.
For each galaxy identified in the simulation
box, we calculate the dust optical depth, extinction curve (based on the grain size
distribution), and intrinsic stellar SED. These quantities together with the dust mass
enable us to estimate the IR luminosity, size of IR-emitting region, dust temperature, 
and observed stellar SED.
We investigate the relations between the output quantities
($L_\mathrm{IR}$--$T_\mathrm{dust}$ and IRX--$\beta_\mathrm{UV}$ relations)
as well as the IR LF.

First, we examine the radius of IR-emitting region for each galaxy in the
simulation. We find that the radius
$2R_\mathrm{dust}$ (where $R_\mathrm{dust}$ is the scale length of dust
distribution)
is consistent with interferometric observations of IR-luminous galaxies.
Remarkably we also reproduce the evolution of IR-emitting radius
toward $z=4$.
Next, we test if our simulation reproduces the 
observed dust-emission properties by \textit{Herschel} at $z=0$.
We confirm that
the IR LF, $L_\mathrm{IR}$--$T_\mathrm{dust}$ relation, and
IRX--$\beta_\mathrm{UV}$ relation at $z=0$ are reproduced by our simulation.

Based on the model `calibrated' by the above observational quantities,
we compare our simulation results with the galaxy properties
of the STUDIES sample at $z\sim 1$--4.  We find that our simulation
reproduces the
$M_*$--$L_\mathrm{IR}$ relation, IR LFs 
(except at the luminous end $L_\mathrm{IR}\sim 10^{13}$ L$_{\sun}$ at $z>2$),
$T_\mathrm{dust}$--$L_\mathrm{IR}$ relation (except the large scatter
seen in the observational sample), and the IRX--$\beta_\mathrm{UV}$ relation
(except some high-$\beta_\mathrm{UV}$ objects in the observational sample).
At $z\gtrsim 3$, we underproduce
the extremely IR-luminous ($\gtrsim 10^{13}$ L$_{\sun}$)
objects even in the higher-resolution simulation. 
This is probably	
due to contributions from sources not included in our simulation such as AGNs. 
For the observed large scatter in the $T_\mathrm{dust}$--$L_\mathrm{IR}$ and
IRX--$\beta_\mathrm{UV}$ relations, 
the variation of dust distribution geometry, which our simulation cannot resolve, could be a possible reason.

The broad success of our simulation indicates that the current dust evolution scenario
is consistent with the observed galaxy properties in deep submm surveys. 
{ It means that the interstellar radiation field, which is roughly determined by
the star formation history, is modelled self-consistently with the dust enrichment history in our cosmological hydrodynamic simulation. 
As for the extinction curve, the SMC-like extinction curve which is expected from our model is consistent with the IR-UV properties of the IR-luminous galaxies. }
In the future, we plan to extend our work using cosmological zoom-in simulations with higher resolution and address the issues that we discussed above.

\section*{Acknowledgments}

We are grateful to C.-Y. Chiang for
useful discussions and comments regarding the AGN fraction in
high-redshift galaxies.
We thank C.-C. Chen, T. Goto, T. Hashimoto, and I. Smail for useful comments.
We acknowledge the STUDIES team and the JCMT for the observational data. 
We are grateful to V. Springel for providing us with the original version
of GADGET-3 code.
Numerical computations were carried out on Cray XC50 at the Center
for Computational Astrophysics,
National Astronomical Observatory of Japan and XL
at the Theoretical Institute for
Advanced Research in Astrophysics (TIARA) in Academia Sinica.
HH thanks the Ministry of Science and Technology for support through grant
MOST 105-2112-M-001-027-MY3 and MOST 107-2923-M-001-003-MY3 (RFBR 18-52-52-006).
CFL, YYC, and WHW are supported by the Ministry of Science and Technology through
grant MOST 105-2112-M-001-029-MY3.
KN and IS acknowledge the support from the JSPS KAKENHI Grant Number JP17H01111.
KN acknowledges the travel support from the Kavli IPMU, World Premier Research Center Initiative (WPI), where part of this work was conducted. 


\appendix 
\section{Stellar SED slope in mixed dust geometry}\label{app:beta_limit}

Here we discuss the behaviour of the IRX--$\beta_\mathrm{UV}$ relation in our
model. If we describe the attenuation of stellar light as $\exp (-\tau_\lambda )$,
the observed luminosities at wavelengths $\lambda_1$ and
$\lambda_2$ are estimated as
\begin{eqnarray} 
L_{\lambda_{1}}&=&\exp(-\tau_{\lambda_{1}})L_{\lambda_{1}}^{0}\\
L_{\lambda_{2}}&=&\exp(-\tau_{\lambda_{2}})L_{\lambda_{2}}^{0}=\exp(-\tau_{\lambda_{2}})L_{\lambda_{1}}^{0}\left( \dfrac{\lambda_{2}}{\lambda_{1}}\right)^{\beta_{\rm UV,int}},
\end{eqnarray}
where $\beta_{\rm UV,int}$ is intrinsic SED slope.
In this case, $\beta_\mathrm{UV}$ is written based on Eq.~(\ref{def_of_beta}) as
\begin{eqnarray} 
\beta_{\rm UV}&=&
\dfrac{\left( \tau_{\lambda_{1}} - \tau_{\lambda_{2}} \right)\log_{10}(\mathrm{e})}
{\log_{10}(\lambda_{2})-\log_{10}(\lambda_{1})}+\beta_{\rm UV, int}.
\label{eq:beta_beta_int}
\end{eqnarray}
If we adopt the mixed geometry, the escape fraction at wavelength $\lambda$ is
written as $f_{\rm esc}(\tau_{\lambda})=[1-\exp(-\tau_{\lambda})]/\tau_{\lambda}$.
If we define the effective optical depth as
\begin{eqnarray}
\tau_{{\rm eff,} \lambda}\equiv -\log_\mathrm{e}\left[\frac{1-\exp(-\tau_{\lambda})}{\tau_{\lambda}}
\right] ,
\end{eqnarray}
we can use equation (\ref{eq:beta_beta_int}) by replacing $\tau_\lambda$ with
$\tau_{\mathrm{eff},\lambda}$.
In the optically thick limit, the effective optical depth becomes 
$\tau_{{\rm eff}, \lambda}\to \log_{e}(\tau_{\lambda})$.
By using this relation, we find that
\begin{eqnarray} 
\left( \beta_{\rm UV}-\beta_{\rm UV,int} \right)\to
\dfrac{\log_\mathrm{10}(\tau_{\lambda_{1}}/\tau_{\lambda_{2}})}
{\log_{10}(\lambda_{2}/\lambda_{1})}
\end{eqnarray}
for the optically thick limit ($\tau_\lambda\gg 1$).
If we adopt the SMC extinction curve, which is appropriate for the massive
galaxies in the simulation, using $\tau_{1650\AA}/\tau_{2500\AA}=1.712$
\citep{2003ApJ...594..279G}, we obtain  the following optically thick limit as
\begin{eqnarray} 
\left( \beta_{\rm UV}-\beta_{\rm UV,int} \right)\to 1.211.
\end{eqnarray}
Therefore, there is an upper limit for $\beta_{\rm UV}-\beta_{\rm UV,int}$
in the mixed geometry. For example, if $\beta_\mathrm{UV,int}=-2.5$, $\beta_\mathrm{UV}$
cannot exceed $-1.3$. In reality, because of contribution from old red stellar population,
higher values of $\beta_\mathrm{UV}$ are also seen in the simulation.

\bibliographystyle{mnras}
\bibliography{ken} 

\bsp	
\label{lastpage}
\end{document}